\documentclass[preprint,12pt]{elsarticle}

\usepackage{amssymb}
\usepackage{amsmath}
\usepackage{upgreek}

\journal{Physica Medica}

\usepackage{hyperref}
\usepackage{tikz}

\hypersetup{ colorlinks,
    citecolor=blue,
    filecolor=blue,
    linkcolor=blue,
    urlcolor=blue
}

\usepackage{xurl}

\usepackage{multirow}

\usepackage{makecell}

\begin{document}

\begin{frontmatter}

\title{Charge collection efficiency of thimble ionization chambers
exposed to ultra-high dose per pulse: experimental and numerical results}

\author[USC]{José Paz-Martín} 
\author[PTB]{Andreas Schüller}
\author[PTB]{Marvin Apel}
\author[USC]{Araceli Gago-Arias}
\author[]{Juan Pardo-Montero$^{\rm c, d}$}
\author[]{Faustino Gómez$^{\rm a, e}$}

\affiliation[USC]{organization={Universidade de Santiago de Compostela},
            addressline={Praza do obradoiro, 0}, 
            city={Santiago de Compostela},
            postcode={15705}, 
            state={Galicia},
            country={Spain}}

\affiliation[PTB]{organization={Physikalisch-Technische Bundesanstalt},
            addressline={Bundesallee 100}, 
            city={Braunschweig},
            postcode={38116},
            country={(Germany)}}
            
\affiliation[IDIS]{organization={Group of Medical Physics and Biomathematics, Instituto de Investigacion Sanitaria de Santiago (IDIS)},
            addressline={Travesía da Choupana}, 
            city={Santiago de Compostela},
            state={A Coruña},
            postcode={15706},
            country={(Spain)}}
            
\affiliation[CHUS]{organization={Department of Medical Physics, Complexo Hospitalario Universitario de Santiago de Compostela},
            addressline={Travesía da Choupana}, 
            city={Santiago de Compostela},
            state={A Coruña},
            postcode={15706},
            country={(Spain)}}

\affiliation[RPL]{organization={Laboratorio de Radiofísica, Universidade de Santiago},
            addressline={Estrada de San Lourenzo}, 
            city={Santiago de Compostela},
            postcode={15782},
            state={A Coruña},
            country={(Spain)}}
            
\begin{abstract}

\noindent {\bf Background:} Commercially available ionization chambers (ICs) exposed to ultra-high dose per pulse (UHDP) exhibit deviations from a linear dose response due to volume recombination. Simulation models have been developed to describe the charge collection efficiency (CCE) but focused on parallel-plate ICs. This study aims to measure and simulate the CCE and polarity effect of thimble ICs in UHDP.\\

\noindent {\bf Methods:} The response of two PinPoint3D T31022 (PP3D) and two PinPoint T31023 (PP) ICs was investigated experimentally at the national metrology institute of Germany (PTB). The ICs were irradiated using the UHDP reference electron beam with dose per pulse up to 9.3 Gy for different voltages. A novel finite element code capable of simulating 1D and 2D geometries was developed.\\

\noindent {\bf Results:} Thimble ICs exhibit a pronounced polarization effect when irradiated with UHDP. When the sign of the collected charge is negative, the CCE is larger than when it is positive. The major contribution to the polarity effect can be attributed to the polarity-dependent charge transport and recombination. Experimental and simulated CCE (polarity effect correction factor) agrees within 1.4 \% (7.0 \%) and 1.6 \% (3.2 \%) for the PP3D and PP ICs, respectively. The CCE of parallel-plate and thimble ICs is related through a geometrical rule.\\

\noindent {\bf Conclusions:} The PP IC shows greater CCE due to its smaller external radius. The numerical model is able to satisfactory reproduce the actual CCE and polarity effect for these two chambers. At UHDP, thimble ICs should be used with caution due to their large polarity effect.

\end{abstract}




\end{frontmatter}



\section{Introduction}

 Commercially available ionization chambers (IC) suffer from volume recombination when exposed to ultra-high dose per pulse~\citep{Petersson_High_2017, Bourgouin_Charge_2023} (DPP) beams, such as those used for FLASH radiotherapy~\citep{Favaudon_Ultrahigh_2014}. In these beams, with DPPs from 0.6 Gy to 10 Gy, the response of the ICs is no longer proportional to the dose due to recombination between opposite charge carriers, compromising their suitability as reference detectors. In the past years, several numerical and phenomenological models have been developed with the aim of describing the response of parallel-plate ICs under ultra-high DPPs~\citep{Petersson_High_2017, Paz-Martin_Numerical_2022, Kranzer_Charge_2022, Gotz_A_2017, Liu_Development_2024, Bancheri_semi-analytical_2024}. These models have shown satisfactory agreement in terms of charge collection efficiency (CCE)~\citep{Kranzer_Charge_2022, Kranzer_Ion_2021,Liu_Evaluation_2024,Subiel_Metrology_2024,Paz-Martin_Numerical_2022} and time-resolved current~\citep{Paz-Martin_Numerical_2022}. For parallel-plate ICs, a 1D model assuming cylindrical symmetry around the axis perpendicular to the electrodes provides a close approximation to the actual geometry. However, the picture is different for thimble ICs, where an idealized 1D cylindrical symmetry is likely not to reproduce the actual behavior of the chamber due to its particular combination of cylindrical and spherical geometries. This is particularly dramatic for ICs where the volume of the spherical tip represents a large fraction of the total volume of the IC. \\

In the literature, the work devoted to volume recombination in cylindrical and spherical
ICs is limited. Recently, for example, an ideal 1D cylindrical simulation code was applied to describe the behavior of a PinPoint T31010 IC for DPPs below 0.4 Gy per pulse \citep{Gotz_Dosimetry_2018}. Fenwick \textit{et al}.~\citep{Fenwick_Collection_2022} developed an analytical model including the free electron fraction for cylindrical IC. In ultra-high DPP, only the response of Nano Razor~\citep{Cavallone_Determination_2022} and Extradin A26 chambers~\citep{Liu_Evaluation_2024} has been experimentally characterized.\\

In this work, the response of two samples of both PinPoint T31023 and PinPoint3D T31022 thimble chambers have been experimentally studied as a function of DPP, pulse duration and bias voltage. Additionally, the time-resolved signals from the ICs were recorded using an oscilloscope. The experimental data is compared to a novel numerical model capable of simulating electric field and charge transport in 1D and 2D geometries using the finite element method.

\section{Materials and methods}

\subsection{Investigated chambers}

This study examines two different models of thimble ICs with similar sensitive volume but different design: the PTW PinPoint 3D T31022 and the PTW PinPoint T31023, hereafter referred to as PP3D and PP, respectively. For each chamber model, two units were assessed—one with internal biasing and the other with external biasing. The key specifications of both IC models are summarized in Table~\ref{ic_parameters}.
Calibration coefficients in terms of absorbed dose to water were determined at PTB’s $^{60}$Co source, using a bias voltage of 400 V. The relative uncertainty of these measurements was 0.4~\% (coverage factor $k = 1$). Although both chambers share similar materials and construction, they differ in their geometric dimensions. \\

\begin{table}[!b]
\begin{center}
\caption{Key parameters of the thimble ICs used in this work. The geometrical information of the chambers were obtained from the manufacturer brochure and the sensitive volume was calculated using the finite element method, including the effect of the guard ring. The calibration coefficients for the ICs in terms of absorbed dose to water were obtained using a bias voltage of 400~V.\label{ic_parameters}\vspace*{2ex}}
\resizebox{\textwidth}{!}{
\begin{tabular} {>{\centering\arraybackslash}m{5cm}cccccc}
\hline
\hline
 & & \multicolumn{2}{c}{\textbf{PinPoint 3D T31022}}  & & \multicolumn{2}{c}{\textbf{PinPoint T31023}} \\
\cline{3-4} \cline{6-7}
 & S/N & 153056 & 152986 & & 170293 & 170313 \\
\hline
Calibration coefficient (Gy\,nC$^{-1}$)  & & 2.587(10) & 2.543(10) & & 2.4267(97) & 2.4542(98)      \\
Inner radius (mm)                        & & \multicolumn{2}{c}{0.30} &  & \multicolumn{2}{c}{0.30} \\
Outer radius (mm)                        & & \multicolumn{2}{c}{1.45} &  & \multicolumn{2}{c}{1.00} \\
Height (mm)                              & & \multicolumn{2}{c}{2.90} &  & \multicolumn{2}{c}{5.00} \\
Sensitive volume (mm$^3$)                          & & \multicolumn{2}{c}{12} &  & \multicolumn{2}{c}{13} \\
\hline
\hline
\end{tabular}}
\end{center}
\end{table}

\begin{figure}[!t]
\vspace{1 cm}
\centering
\includegraphics[width = 13.5 cm]{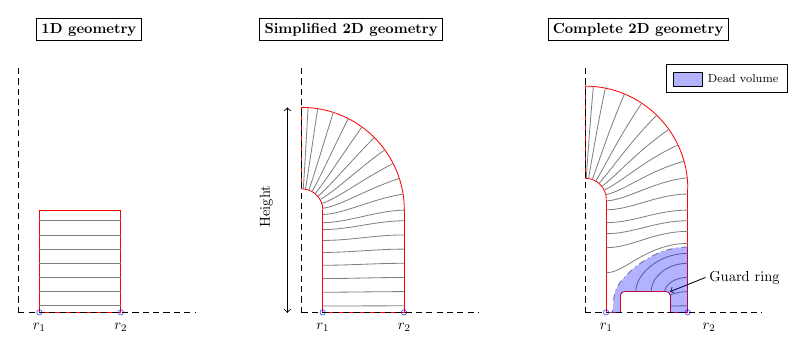}
\caption{Simplified diagrams of the different IC geometries considered in this study. $r_1$ corresponds to the inner radius and $r_2$ to the outer radius of the IC cavity. All the geometries assume a cylindrical symmetry around the vertical axis. In the complete 2D geometry the guard ring is assumed to be a perfect conductor. Gray lines shows the electric field lines and the dead volume is highlighted in blue.}
\label{Fig1}
\end{figure}

Figure~\ref{Fig1} illustrates the geometric approaches used in the simulations (not to scale). In this work, the term '1D cylindrical (spherical) model' refers to the solution of transport equations depending solely on the radial coordinate, assuming a purely cylindrical (or spherical) geometry. In contrast, 2D models account for variations along both the radial and axial coordinates of the chamber, incorporating structural features such as the chamber tip and guard ring.\\

For the 1D and simplified 2D models, the physical volume of the IC was employed to calculate the released charge. In contrast, simulations using the full 2D geometry --which includes the guard ring electrode, accounting for the IC's dead volume \citep{Delfs_Role_2021, Pojtinger_Finite_2019}-- utilized the actual sensitive volume of the IC, determined using the simulated electric field lines ending in the central electrode.

\subsection{Experimental setup}

The experimental characterization of the ICs was conducted at the metrological electron accelerator facility of the German national metrology institute, the Physikalisch-Technische Bundesanstalt (PTB) using the ultra-high pulse dose rate reference electron beam~\citep{Schüller_The_2018, Bourgouin_Characterization_2022}.\\

The ICs were irradiated using an electron beam with an energy of 20 MeV. A wide range of DPPs was achieved using different combination of source-to-surface distance (SSD) and aluminum scattering foils as well as a variable diaphragm at the beginning of the beamline to reduce the charge per pulse. Overall, 5 different setups were used: 70~cm SSD (without and with 1~mm Al) and 90 cm SSD (without and with, 1~mm, 2~mm and 6~mm of Al). Two pulse durations were considered in this investigation, 1.0~$\upmu$s and 1.9~$\upmu$s. The chambers were positioned in a PMMA water phantom at a fixed water equivalent depth of 45.2~mm, close to the reference depth of the different setups. The reference dosimetry was performed using a flashDiamond~\citep{Marinelli_Design_2022, Kranzer_Response_2022} (PTW, Freiburg, Germany), calibrated by means of PTB’s alanine dosimetry system~\citep{Bourgouin_Absorbed_2022, Vörös_Relative_2012,Subiel_Metrology_2024}. Pulse-to-pulse deviations produced by variations in the beam charge per pulse were corrected using the signal of an in-flange beam current transformer (ICT)~\citep{Schüller_Traceable_2017} (Bergoz, Saint Genis Pouilly, France). The ICT was calibrated with the flashDiamond for each setup, pulse duration and  diaphragm aperture to determine the reference DPP. The bias voltage was supplied to the ICs using an in-house high voltage source, varying from 200~V up to 500~V in steps of 100~V for both positive and negative polarity. The charge per pulse from the ICs was measured using a Keythley 616 electrometer with a 33~nF capacitor placed between the core wire and the inner shield of the triaxial cable. This capacitor prevents a non-linear response of the electrometer due to a large signal coming from the detector, as reported by several authors~\citep{Bourgouin_Charge_2023, Marinelli_Design_2022, Gómez_Development_2022}.\\

To measure the time-resolved current from the ICs, the collecting electrode was connected to a HVA-200M-40-B voltage amplifier (FEMTO, Berlin, Germany), while the bias voltage was applied to the external electrode of the IC. Simultaneously, the beam pulse structure was monitored using the ICT. Both the IC and ICT currents were recorded simultaneously using the two input channels of a fast 14 bit digitizer (Spectrum M3i.4142), following a procedure similar to that described by Paz-Martín \textit{et al.}~\citep{Paz-Martin_Numerical_2022}. Time-resolved measurements were performed using bias voltages ranging from 75~V to 500~V.

\subsection[CCE and $k_{\rm pol}$ determination]{Charge collection efficiency and the polarity effect correction factor determination}

The averaged CCE, denoted as $f$ in the equations, was determined experimentally as the average between the positive and the negative CCE ($f_{\pm}$) at the same DPP. $f_{\pm}$ was determined as the ratio between the DPP obtained with the ICs (without applying any correction for volume recombination and polarity) over the reference DPP ($D_{\rm ref}$):
\begin{equation}
f = \frac{f_+ + f_-}{2} \quad \quad f_{\pm} = \frac{|\,Q_{\pm}|\, N_{\rm D,w,Q_0} \, k_{\rm elec} \, k_{\rm TP} \, k_{\rm Q,Q_0}}{D_{\rm ref}}
\end{equation}
where $Q_+$ and $Q_-$ are the collected charge when its sign is positive or negative, respectively.\\

The beam quality correction factors ($k_{\rm Q,Q_0}$) were calculated using the EGSnrc Monte Carlo code system~\cite{Kawrakow_EGSnrc_2000} (version 4.0, 2023a). The linear accelerator beam line and exit window were modeled using the \texttt{BEAMnrc}~\citep{Rogers_BEAM_1995} user code, adopting Bourgouin \textit{et al.}~\citep{Bourgouin_Characterization_2022} approach. To compute the ICs’ perturbation factors, the \texttt{egs\_chamber}~\citep{Wulff_Efficiency_2008} user code was used, implementing the detailed geometries provided by the ICs’ manufacturer (PTW). The $^{60}$Co calibration beam was modeled as a square 10~cm$~\times$~10~cm divergent beam with a source-to-surface distance of 100~cm and the same energy fluence spectrum as Mora \textit{et al.}~\citep{Mora_Monte_1999}. Both electrons and photons were transported until their kinetic energy fell below 1~keV. For the reference electron beam simulations, a cut-off of 10~keV for both electrons and photons was used, as no significant effect was previously reported~\citep{Muir_Monte_2014}. All the simulated $k_{\rm Q,Q_0}$ were obtained with a relative statistical uncertainty below 0.35~\%. Additionally the $k_{\rm Q,Q_0}$ factors were experimentally determined for two different detectors of each type by cross-calibrating them with a PTW 34001 Roos chamber in an electron beam with a nominal Energy of 20 MeV (R$_{50}$ = 77.4 mm) produced by a clinical LINAC (Elekta Synergy).

\noindent The polarity effect correction factor ($k_{\rm pol}$) is defined as:

\begin{equation}
k_{\rm pol} = \frac{|Q_+| + |Q_-|}{2\,|Q_+|}
\end{equation}
analogous to TRS-398~\citep{IAEA_Absorbed_2024} and TG-51~\citep{Almond_AAPM_1999} dosimetry international protocols.\\

The uncertainty of the experimentally determined CCE is estimated to be between 1.2~\% and 1.8~\% (for $k$ = 1), similar to that reported in previous studies conducted using a comparable setup~\citep{Bourgouin_Charge_2023, Paz-Martin_Numerical_2022}.

\subsection{The numerical model}
The transport equations that describe the drift of the charge carriers inside the ICs were adapted from Paz-Martín \textit{et al.}~\citep{Paz-Martin_Numerical_2022} to describe geometries with cylindrical and spherical symmetry. The system of differential equations considered are: 
\begin{align}
\frac{\partial n_+(\boldsymbol{x}, t)}{\partial t} =&\;n_0(\boldsymbol{x}, t) - \alpha\; n_+(\boldsymbol{x}, t)\; n_-(\boldsymbol{x}, t) - \theta\; n_+(\boldsymbol{x}, t)\; n_e(\boldsymbol{x}, t) \nonumber \\  & -\nabla\cdot\left[\boldsymbol{E}(\boldsymbol{x}, t)\; \mu_+ \; n_+(\boldsymbol{x}, t)\right] + \eta(|\boldsymbol{E}|, t)n_e(\boldsymbol{x},t) + \nabla\cdot\left[D_+\nabla n_+(\boldsymbol{x}, t)\right], \nonumber \\
\frac{\partial n_-(\boldsymbol{x}, t)}{\partial t} = &~\gamma(|\boldsymbol{E}|, t) \; n_e(\boldsymbol{x}, t) -\alpha\; n_+(\boldsymbol{x}, t) \; n_-(\boldsymbol{x}, t) \label{ICtransport} \\ & + \nabla\cdot\left[\boldsymbol{E}(\boldsymbol{x}, t)\; \mu_- \; n_-(\boldsymbol{x}, t)\right] + \nabla\cdot\left[D_-\nabla n_-(\boldsymbol{x}, t)\right], \nonumber \\
\frac{\partial n_e(\boldsymbol{x}, t)}{\partial t} =&\;n_0(\boldsymbol{x}, t)-\gamma(|\boldsymbol{E}|, t) \; n_e(\boldsymbol{x}, t) - \theta\; n_+(\boldsymbol{x}, t)\; n_e(\boldsymbol{x}, t) \nonumber \\ & + \nabla\cdot\left[\boldsymbol{v_e}(\boldsymbol{x}, t)\; n_e(\boldsymbol{x}, t)\right] + \eta(|\boldsymbol{E}|, t)n_e(\boldsymbol{x},t) + \nabla\cdot\left[D_e(\boldsymbol{E}, t)\nabla n_e(\boldsymbol{x}, t)\right], \nonumber
\end{align}
being $\boldsymbol{x}$ and $t$ the 3D position vector and time, respectively.
All the symbols used are defined in Table~\ref{tableSymbols}. As boundary conditions, the charge carrier densities are set to zero on the electrode of the same polarity. The electric field perturbation is included by solving the Poisson equation for each time step of the simulation:
\begin{equation}
\nabla\cdot\boldsymbol{E}(\boldsymbol{x}, t) = \frac{e}{\varepsilon} \left[n_+(\boldsymbol{x}, t) - n_-(\boldsymbol{x}, t) - n_e(\boldsymbol{x}, t)\right],
\label{eField}
\end{equation}
\begin{table}[!t]
\caption{Definition of the symbols used in Equation~\ref{ICtransport} and~\ref{eField}.}
\vspace{0.5 cm}
\centering
\resizebox{\textwidth}{!}{
\begin{tabular}{>{\centering\arraybackslash}m{4 cm}||c||>{\centering\arraybackslash}m{6 cm}}
\hline
\hline
\multicolumn{1}{c||}{\textbf{\small{Symbol}}} &  \multicolumn{1}{c||}{\textbf{\small{Units}}} & \multicolumn{1}{c}{\textbf{\small{Definition}}} \\ \hline
$n_+(\boldsymbol{x}, t)$, $n_-(\boldsymbol{x}, t)$, $n_e(\boldsymbol{x}, t)$  & m$^{-3}$        & \small{Positive, negative and electron densities, respectively} \\ \hline
$n_0(\boldsymbol{x}, t)$ & m$^{-3}$ s$^{-1}$ & \small{Charge released per unit of time and volume that escapes initial recombination} \\ \hline
$\boldsymbol{E}(\boldsymbol{x}, t)$ & V m$^{-1}$  & \small{Electric field} \\ \hline
$\gamma(|\boldsymbol{E}|, t)$     		  & s$^{-1}$  & \small{Electron attachment rate} \\ \hline
$\eta(|\boldsymbol{E}|, t)$     		  & s$^{-1}$  & \small{Electron multiplication coefficient} \\ \hline
$\alpha$     		  & m$^3\;$s$^{-1}$  & \small{Ion-ion recombination coefficient} \\ \hline
$\theta$     		  & m$^3\;$s$^{-1}$  & \small{Electron-ion recombination coefficient} \\ \hline
$\boldsymbol{v_e}(\boldsymbol{E}, t)$     		  & m$\;$s$^{-1}$  & \small{Electron drift velocity} \\ \hline
$\mu_+$, $\mu_-$      & m$^2\;$V$^{-1}\;$s$^{-1}$  & \small{Positive and negative ion mobility, respectively} \\ \hline
$e$                   & C         & \small{Elementary charge} \\ \hline
$\varepsilon$         & C V$^{-1}$ m$^{-1}$  & \small{Air permittivity} \\
\hline
\hline
\end{tabular}}
\label{tableSymbols}
\end{table}
In addition, for any two points $P_1$, $P_2$ in different electrodes connected by a path $\boldsymbol{l}$, the integral of the electric field must be equal to the voltage difference $U$ between these two electrodes:
\begin{equation}
-\int_{P_1}^{P_2}\boldsymbol{E}(\boldsymbol{x}, t)\cdot d\boldsymbol{l} = U	\quad \forall t
\end{equation}
We have expressed this set of equations in vector form to enhance generality. The expressions for the gradient and divergence operators will depend on the chosen curvilinear coordinate system.\\

To develop a general code capable of simulating a wide variety of geometries, the finite element method was used instead of the traditional finite difference scheme employed for parallel-plate ICs. The code was implemented using the Python package \texttt{dolfinx} \citep{Baratta_DOLFINx_2023} (version 0.9.0). To generate the 2D meshes from the IC geometries, the \texttt{GMSH} program \citep{Geuzaine_Gmsh_2009} was used. The temporal discretization of the equations was performed using an adaptive second-order backward differentiation scheme \citep{Celaya_Implementation_2014}.

\begin{table}[!t]
\begin{center}
\vspace*{2ex}
\caption{Electron and ion transport parameters used this work.\label{traspParameters}}
\vspace{0.5 cm}
\begin{tabular} {>{\centering\arraybackslash}m{4.5 cm}||>{\centering\arraybackslash}m{8 cm}}
\hline
\hline
\textbf{Transport parameter} & \textbf{Description} \\
\hline
Electron velocity & Calculated using \texttt{Magboltz}~\citep{Biagi_Monte_1999} \\
Electron attachment rate & Calculated using \texttt{Magboltz}~\citep{Biagi_Monte_1999} \\
Ion mobilities & From Zhang \textit{et al.}~\citep{Zhang_Prediction_2019} and Boissonnat \textit{et al.}~\citep{Boissonnat_Measurement_2016} \\
Ion-ion recombination coefficient & Depending on the ion mobilities: 
$0.8 \times 10^{-12} \,\text{m}^3 \,\text{s}^{-1}$ for Zhang \textit{et al.}~\citep{Zhang_Prediction_2019} and $1.4 \times 10^{-12} \,\text{m}^3 \,\text{s}^{-1}$ for Boissonnat \textit{et al.}~\citep{Boissonnat_Measurement_2016}, respectively. \\
Electron-ion recombination coefficient & $4.45 \times 10^{-12} \,\text{m}^3 \,\text{s}^{-1}$ from Gotz \textit{et al.}~\citep{Gotz_A_2017} (when using Boissonnat \textit{et al}. mobilities). \\
Longitudinal diffusion coefficient & Calculated using \texttt{Magboltz}~\citep{Biagi_Monte_1999} for electrons and Nernst-Townsend relation for positive and negative ions. \\
Electron multiplication coefficient & Calculated using \texttt{Magboltz}~\citep{Biagi_Monte_1999} \\
Temporal structure & Pulsed beam with rectangular shape. To compare time-resolved signals, the actual beam current was introduced.\\
Spatial discretization & Variable, typically below 40~$\upmu$m. \\
Time discretization & Adaptative time step. \\
\hline
\hline
\end{tabular}
\end{center}
\end{table}

The time-resolved signal produced by each charge carrier drifting in the IC (denoted as $I_i(t)$, being $i$ positive ions, negative ions, or electrons) was calculated using the Shockley–Ramo theorem \citep{Ramo_Currents_1939}:
\begin{equation}
I_{i}(t) = e \int \,n_{i}(\boldsymbol{x}, t)\,\boldsymbol{E_w}(\boldsymbol{x})\cdot \boldsymbol{v_i(\boldsymbol{E}, t)}\,d\boldsymbol{x}
\label{ind}
\end{equation}

\noindent where $\boldsymbol{v_i}$  denotes the actual drift velocity of the charge carriers, and $\boldsymbol{E_w}(\boldsymbol{x})$ represents the weighting electric field. This field is computed under the following conditions: all free charges are removed from the volume, the collecting electrode set to 1~V and rest of the electrodes grounded. 
To enable comparison between experimental and simulated time-resolved current signals, the actual beam pulse structure was incorporated into the simulation. The electron and ion transport parameters employed in this study are listed in Table~\ref{traspParameters}.\\

Based on the convergence behavior observed when using progressively smaller time steps, the uncertainty of the numerical model --arising from both temporal and spatial discretizations-- has been estimated to be below 1~\% in terms of CCE.

\subsection{The Fenwick \textit{et al.} analytical model}
Recently, Fenwick \textit{et al.} extended the analytical model developed for calculating the CCE of parallel-plate ICs~\citep{Fenwick_Collection_2022}, in the absence of electric field perturbation, to describe the CCE of cylindrical ICs~\citep{Fenwick_Collection_2024}. In their model, the CCE ($f_{\rm fk}$) can be written as:
\begin{equation}
f_{\rm fk} = \frac{1}{u}\ln\left[1 + u\;h(u, \Delta)\right], \quad\quad u = \frac{\alpha\, n_0\, g}{(\mu_++\mu_-)\,U}, \quad\quad \Delta=\frac{\gamma\,g}{\mu_e\,U}
\label{fenwick}
\end{equation}
where
\begin{equation}
h(u, \Delta) = \frac{1}{\Delta}\exp\left(\frac{u}{\Delta}\right)\left[{\rm E}_1\left(\frac{u}{\Delta}\exp(-\Delta)\right)-{\rm E}_1\left(\frac{u}{\Delta}\right)\right],
\end{equation}
$g$ is a parameter that depends on the chamber geometry, $\mu_e$ is the electron mobility and E$_1$ the standard exponential integral function:
\begin{equation}
{\rm E}_1(x) = \int_x^\infty\frac{e^{-y}}{y}\,dy
\end{equation}
The parameter $g$ has been described previously in Boag's recombination models~\citep{Boag_Ionization_1950}. For a parallel-plate IC with a distance between electrodes $d$, the geometrical parameter $g$ is:
\begin{equation}
g_{\rm pp} = d^2
\end{equation}
while for a cylindrical IC with a collector radius $r_1$ and a cavity radius $r_2$,
\begin{equation}
g_{\rm cyl} = \frac{(r_2^2-r_1^2)}{2}\ln\left(\frac{r_2}{r_1}\right)
\end{equation}

\section{Results}

\subsection{Beam quality correction factors}

The experimental and calculated beam quality correction factors ($k_{\rm Q,Q_0}$) for the PP3D and PP ICs are shown in Table~\ref{kQs}. The observed variation in $k_{\rm Q,Q_0}$ across different setups is consistent with changes in the water-to-air stopping power ratio, which result from the electron energy degradation and straggling with aluminum scattering foils of different thickness~\citep{Burns_R50_1996}. The 20~MeV Elekta Synergy beam is comparable in terms of beam quality to the 90~cm~+~2~mm~Al setup, with an $R_{50}$ value of 77.1~mm. The experimental determined $k_{\rm Q,Q_0}$, averaged between the two ICs samples, show a good agreement with the simulated values.

\begin{table}[!h]
\begin{center}
\caption{Computed $k_{\rm Q,Q_0}$ for the PinPoint T31023 and the PinPoint 3D T31022 ICs for the different setups used at the PTB’s ultra-high pulse dose rate reference electron beam and experimental $k_{\rm Q,Q_0}$ determined in the 20~MeV Elekta Synergy. All the calculations were performed using the \texttt{EGSnrc} Monte Carlo system. The statistical uncertainty of the calculated values is always below 0.35~\% (coverage factor $k = 1$) and the uncertainty of the experimental $k_{\rm Q,Q_0}$ are estimated to be 0.85~\%. 
\label{kQs}}
\vspace{0.5 cm}
\begin{tabular} {>{\centering\arraybackslash}m{4cm}||cc}
\hline
\hline
\textbf{Setup}        & \textbf{PinPoint3D T31022} & \textbf{PinPoint T31023} \\
\hline
70~cm                 &  0.9082                   & 0.9091                     \\
70~cm~+~1~mm~Al       &  0.9089                   & 0.9086                     \\
90~cm                 &  0.9070                   & 0.9056                     \\
90~cm~+~1~mm~Al       &  0.9098                   & 0.9054                     \\
90~cm~+~2~mm~Al       &  0.9116                   & 0.9121                     \\
90~cm~+~6~mm~Al       &  0.9226                   & 0.9268                     \\
20 MeV Elekta Synergy & 0.9145                     & 0.9150                    \\
\hline
\hline
\end{tabular}
\end{center}
\end{table}

\subsection{Comparison between numerical and analytical model}

\begin{figure}[!b]
\centering
\includegraphics[width = 8 cm]{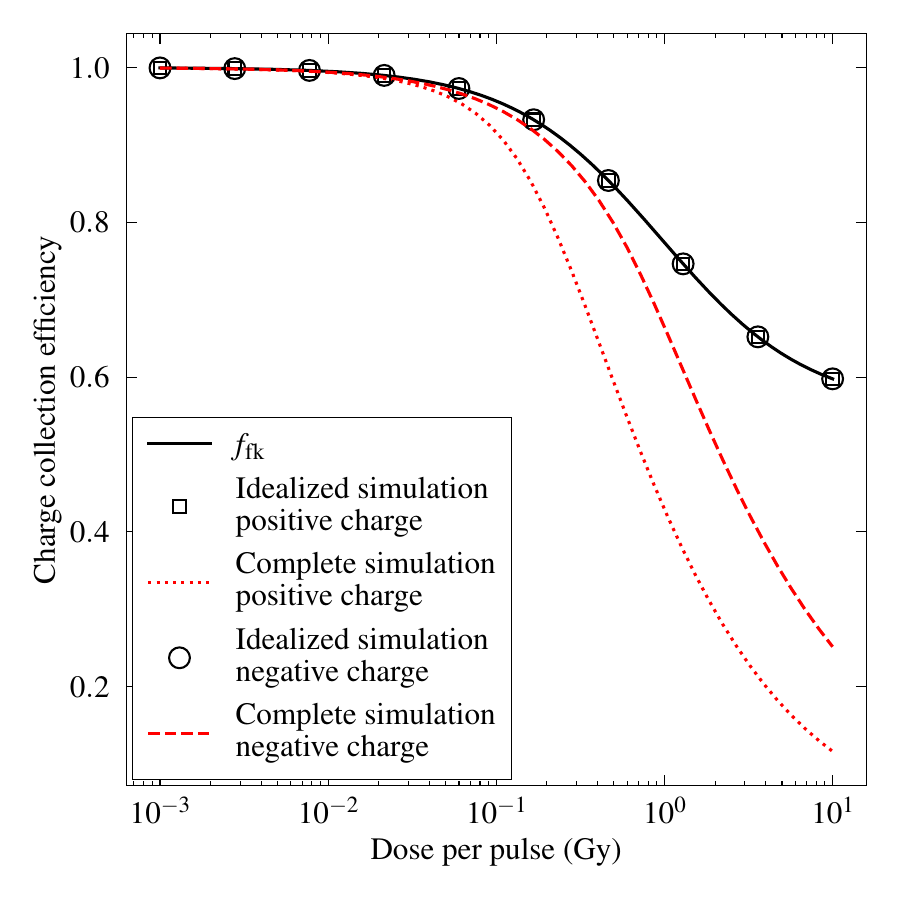}
\vspace{-0.6 cm}
\caption{Comparison between the CCE calculated using the numerical model and the analytical model of Fenwick \textit{et al.}~\citep{Fenwick_Collection_2024} as a function of the DPP for a PinPoint3D T31022 IC to illustrate the polarity and electric field perturbation effects. For this calculation the chamber is idealized as a tip-less cylindrical geometry biased at 300~V. The simulation assumes: i) no electric field perturbation, ii) charge released instantaneously in the active volume (zero pulse duration) and iii) electrons are collected immediately after the release of the charge. Positive (negative) charge corresponds to positive (negative) polarity using external polarization.}
\label{Fig2}
\end{figure}
To initially validate the developed numerical model, the simulated CCE of an idealized cylindrical 1D IC with similar geometrical properties to the PP3D IC was compared to the results of Fenwick~\textit{et al.}~\citep{Fenwick_Collection_2024} analytical model (Equation~\ref{fenwick}). For this comparison, the same transport parameters and simplifications considered by Fenwick \textit{et al.}~\citep{Fenwick_Collection_2024} were incorporated in the numerical model, namely: electric field perturbation was disabled, charge was released instantaneously within the active volume (zero pulse duration), and electron drift was assumed to occur instantaneously after the release of the charge. Figure~\ref{Fig2} shows the CCE obtained from the numerical and analytical models as a function of the DPP (from 1~mGy to 10~Gy per pulse) using a bias voltage of 300~V. Discrepancies between the simplified numerical simulation and the analytical values were always below 0.01~\% for the whole range of DPPs. For completeness, the results of the complete numerical simulation enabling all of the aforementioned effects are also shown.\\

The complete simulation exhibits a different positive and negative CCE depending on the applied bias voltage polarity, showing a greater negative CCE (i.e. negative charge drifting to the central electrode) at large DPP. The difference between positive and negative polarity can amount to up to 50~\% at the maximum simulated DPP of 10~Gy per pulse. This effect results in an additional contribution to the polarity effect for cylindrical ICs, arising from the different volume recombination inside the IC depending on the polarity of the bias voltage that becomes the main contribution to the polarity effect at large DPPs.\\

As highlighted by Fenwick \textit{et al.}~\citep{Fenwick_Collection_2024}, when the electron attachment coefficient and the electron mobility are assumed to be independent of the electric field strength and electric field perturbation is neglected, no difference in CCE between positive and negative polarities is expected. The observed polarity effect can be attributed to a variation in the free electron fraction, which depends on the bias voltage polarity. This is a consequence of the dependence of electron transport parameters on the electric field, combined with the presence of electric field perturbation. As a result, CCE is higher when electrons drift toward the central electrode of the chamber. This corresponds to negative sign collected charge, achieved either by applying negative polarity to the external electrode or positive polarity to the collecting electrode.\\

Unless explicitly stated otherwise in the rest of the text, CCE refers to $f$, the average value obtained from measurements under both positive and negative polarity of the IC.

\subsection{Polarity effect in cylindrical ionization chambers}

For a better comprehension of the polarity effect arising from recombination in cylindrical ICs, charge transport was studied in a 1D cylindrical IC with geometrical characteristics similar to those of the PP3D IC. Numerical simulations were performed using a DPP of 3~Gy, a pulse duration of 1.9~$\upmu$s and a bias voltage of 300~V for both positive and negative polarity. These results were obtained with the transport parameters listed in Table~\ref{traspParameters}. 

\begin{figure}[!t]
    \centering
    \includegraphics[width=8 cm]{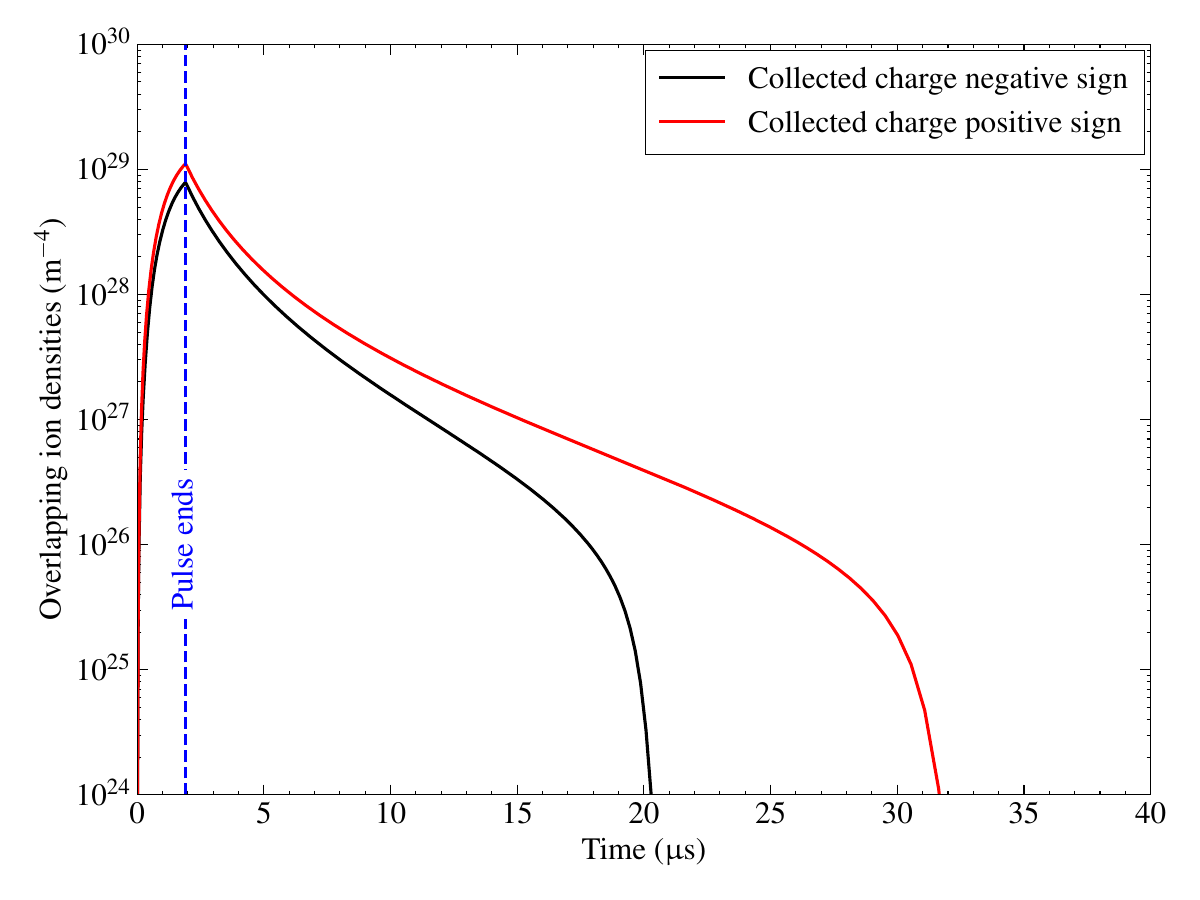}
    \caption{Simulated overlap of ion densities ($\int_{r_1}^{r_2}n_+(r, t)\,n_-(r, t)\,r\,dr$) over time for a 3~Gy per pulse beam delivery with a pulse duration of 1.9~$\upmu$s and a chamber with similar geometrical properties to those of the PP3D IC biased with $\pm$~300~V.}
    \label{FigProv2}
\end{figure}

It is important to emphasize that the dominant contribution to charge loss comes from ion-ion recombination, as electron--ion recombination is a subdominant process. Ion-ion recombination occurs through the overlap of positive and negative ion densities within the active volume of the chamber, quantified by $\int_{r_1}^{r_2}n_+(r, t)\,n_-(r, t)\, r \,dr$ whose time evolution is shown in Figure~\ref{FigProv2} for both polarities.
The overlap between the negative and positive ion distributions reaches its maximum value at the end of the radiation pulse. As clearly illustrated in Figure~\ref{FigProv2} the ion density overlap --and therefore the recombination effect-- is greater when the collected charge at the central electrode is positive.

\begin{figure}[!b]
    \centering
    \includegraphics[width=13.5 cm]{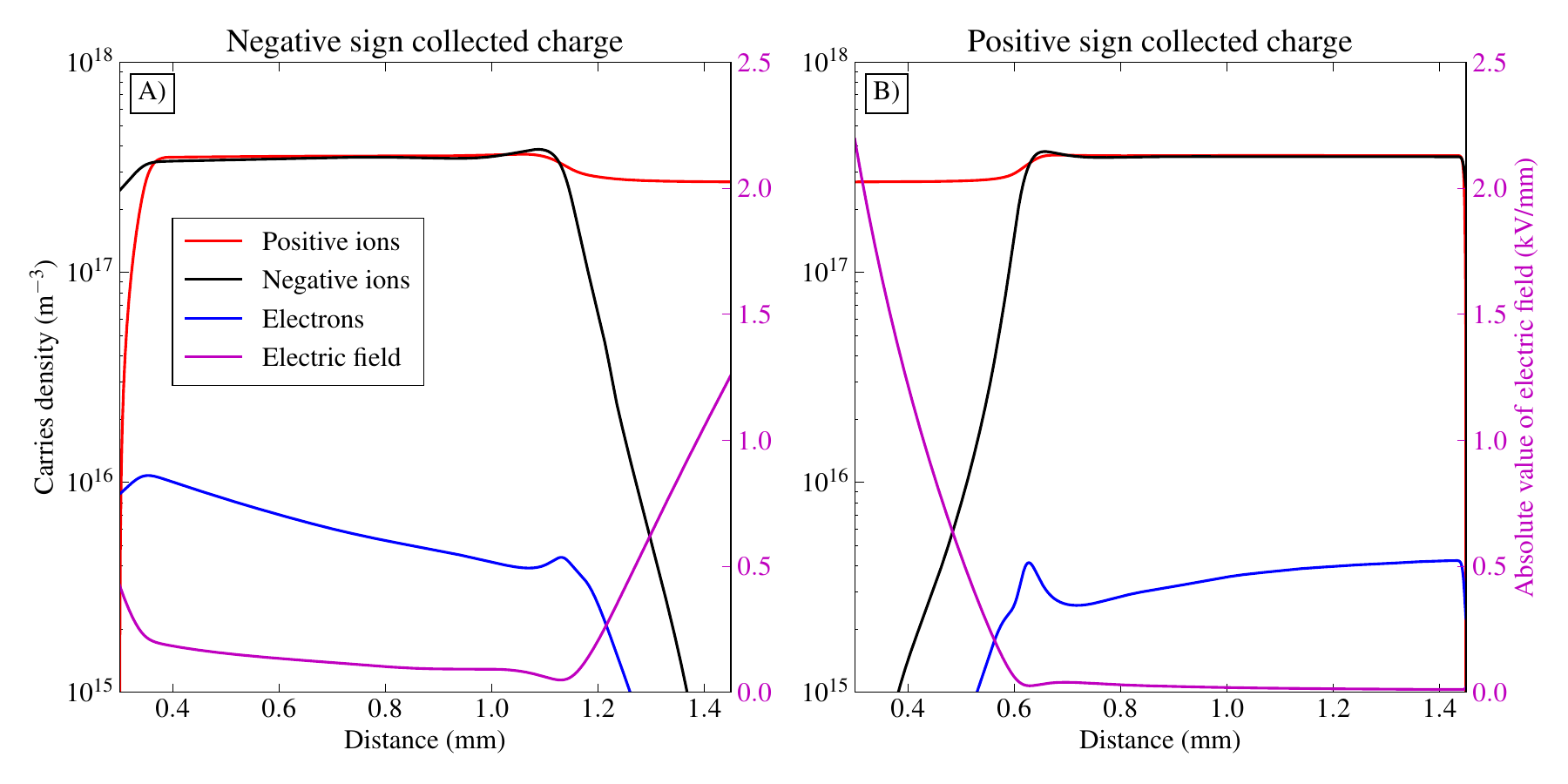}
    \caption{Simulated electric field and charge carrier densities as a function of the radial coordinate for a 1D cylindrical ionization chamber with similar geometrical properties to those of the PP3D IC  at the end of a 1.9~$\upmu$s pulse. Simulations correspond to a positive and negative bias voltage of 300~V and a DPP of 3~Gy.}
    \label{FigProv1}
\end{figure}

The simulated charge carrier densities and electric field as a function of the radial coordinate are shown in 
Figure~\ref{FigProv1} at the end of the radiation pulse.  When the collected charge is negative (panel A), the main region of ion overlap is located closer to the central electrode, whereas it shifts toward the external electrode (panel B) when the collected charge is positive. In Figure~\ref{FigProv1} the electric field perturbation caused by net charge imbalance leads to a significantly reduced field value in the ion overlap region after the pulse delivery. Nevertheless, the higher electric field magnitude in the overlap region for negative polarity (panel A) has a twofold effect: it produces a faster drift of the charges --thereby reducing both the ion-overlap time and the electron transit time-- and it increases the free electron fraction, which again reduces charge loss  by depleting the negative ion population. This second effect is enhanced by the increase in electron lifetime in air with electric field magnitude at intermediate values (0.1~kV/mm to 1.0~kV/mm), resulting in a lower amount of negative ions when electrons drift toward the central electrode. These effects contribute to a higher recombination in a thimble chamber when the polarity of the collected charge is positive. The same physical principles apply for lower DPP deliveries, leading to a higher CCE that still retains a polarity dependence as shown in Figure~\ref{Fig3}. 

\subsection[Weighted cylindrical and spherical geometry]{Description of thimble ICs as weighted cylindrical and spherical geometry}
\label{weighted_CCE}

The impact of the IC geometry on the CCE was evaluated by comparing the results using the geometries from Figure~\ref{Fig1}: a 1D cylindrical geometry ($f_{\text{cyl}}$), a 1D spherical geometry ($f_{\text{sph}}$), a volume-weighted average model ($f_{w}$), a simplified 2D geometry ($f_{\text{2D}}$), and the complete 2D geometry ($f_{\text{2D+gR}}$). The volume-weighted average model is defined as the positive and negative CCE ($f_{\pm w}$) calculated using the volume covered by the spherical tip ($V_{\rm sph}$) and the cylindrical part ($V_{\rm cyl}$):
\begin{equation}
f_{\pm \rm w} = w f_{\pm \rm cyl} + (1 - w)  f_{\pm\rm sph} \quad\quad w = \frac{V_{\rm cyl}}{V_{\rm tot}}
\end{equation}

\noindent
where $V_{\text{tot}} = V_{\text{sph}} + V_{\text{cyl}}$ is the total physical volume of the IC.\\

\begin{figure}[!t]
\centering
\includegraphics[width = 12.0 cm]{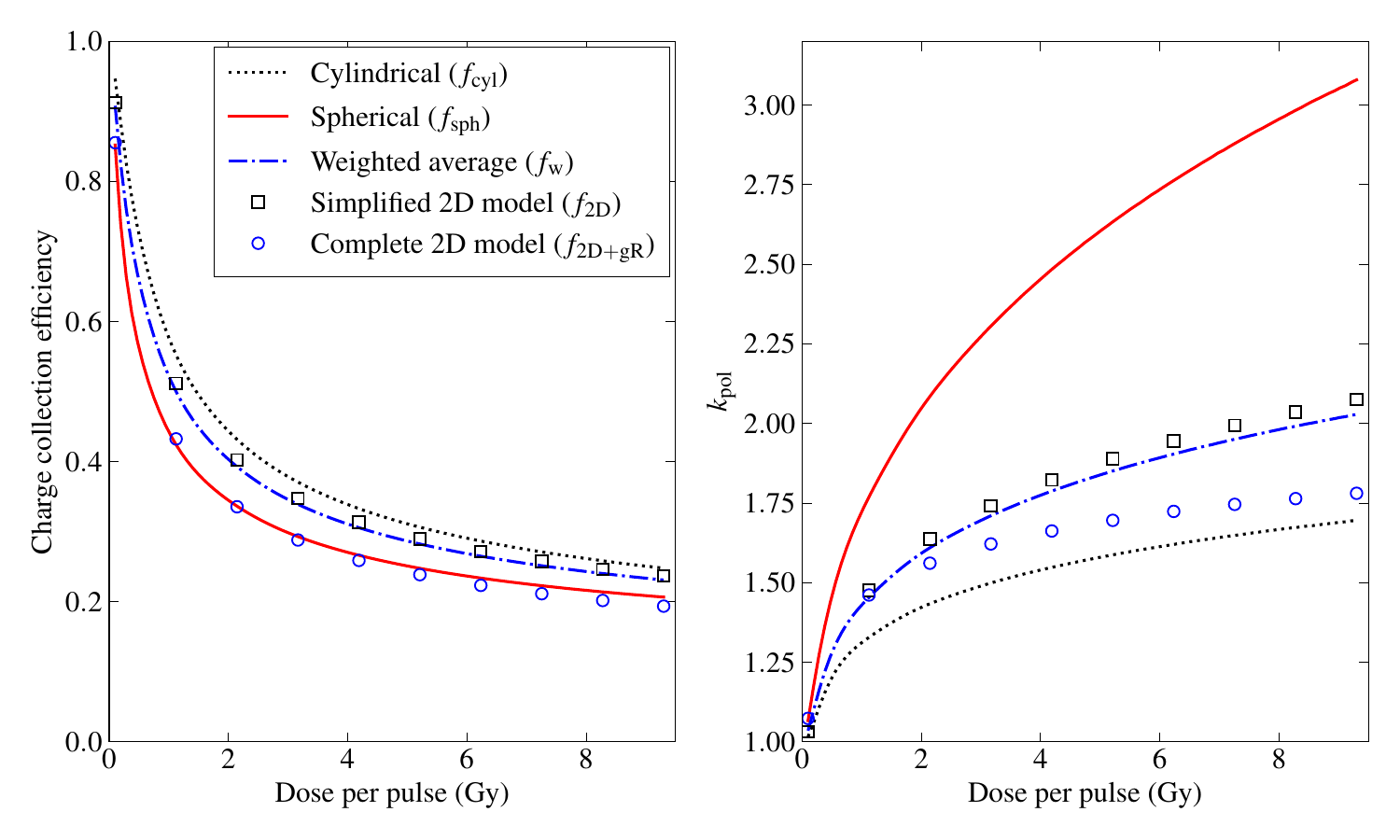}
\caption{Comparison of the CCE and $k_{\rm pol}$ simulations using a 1D cylindrical geometry ($f_{\rm cyl}$), a 1D spherical geometry ($f_{\rm sph}$), a volume-weighted average model ($f_{\rm w}$) a simplified 2D geometry ($f_{\rm 2D}$) and the complete 2D geometry with guard ring ($f_{\rm 2D+gR}$) for a PinPoint3D T31022 IC as a function of the DPP. Calculations were conducted using a pulse duration of 1.9~$\upmu$s, a bias voltage of 300~V and ion mobilities from Boissonnat \textit{et al.}~\citep{Boissonnat_Measurement_2016}}
\label{Fig3}
\end{figure}

Figure~\ref{Fig3} shows the results for a PP3D IC as a function of the DPP (from 0.1~Gy to 9.3~Gy per pulse) using a bias voltage of 300~V and a pulse duration of 1.9~$\upmu$s. The maximum discrepancy between the volume-weighted average and the simplified 2D model in terms of CCE is 2.5~\% and 2.3~\% for $k_{\rm pol}$. When the guard ring is considered in the simulation, the CCE and $k_{\rm pol}$ deviate from the simplified 2D model by up to 18~\% and 14~\%, respectively, at the maximum DPP. It is important to note that accounting for the guard ring also affects the effective volume of the IC, and consequently the actual collected charge. The change in chamber volume contributes to approximately 7~\% of the variation in CCE. Furthermore, when the electric field is perturbed, the chamber sensitive volume may also be altered.\\

In contrast to the PP3D IC, the influence of the selected simulation geometry for the PP IC is less significant. For this chamber, the maximum discrepancy between the volume-weighted average and the simplified 2D geometry is 0.2~\% for CCE and 0.2~\% for $k_{\rm pol}$. The maximum difference between the simplified 2D geometry and the complete 2D geometry reaches 1.7~\% for CCE and 1.9~\% for $k_{\rm pol}$. For both ICs, the presence of a guard ring leads to a reduction in both CCE and $k_{\rm pol}$. Unless explicitly stated, the following sections use the complete 2D geometry for the numerical simulations.

\vspace{-0.2 cm}

\subsection[Experimental versus simulated CCE and $k_{\rm pol}$]{Comparison between experimental and simulated charge collection efficiency and polarity effect}
\label{CCE_kpol}

\begin{figure}[!h]
\centering
\includegraphics[width = 6 cm]{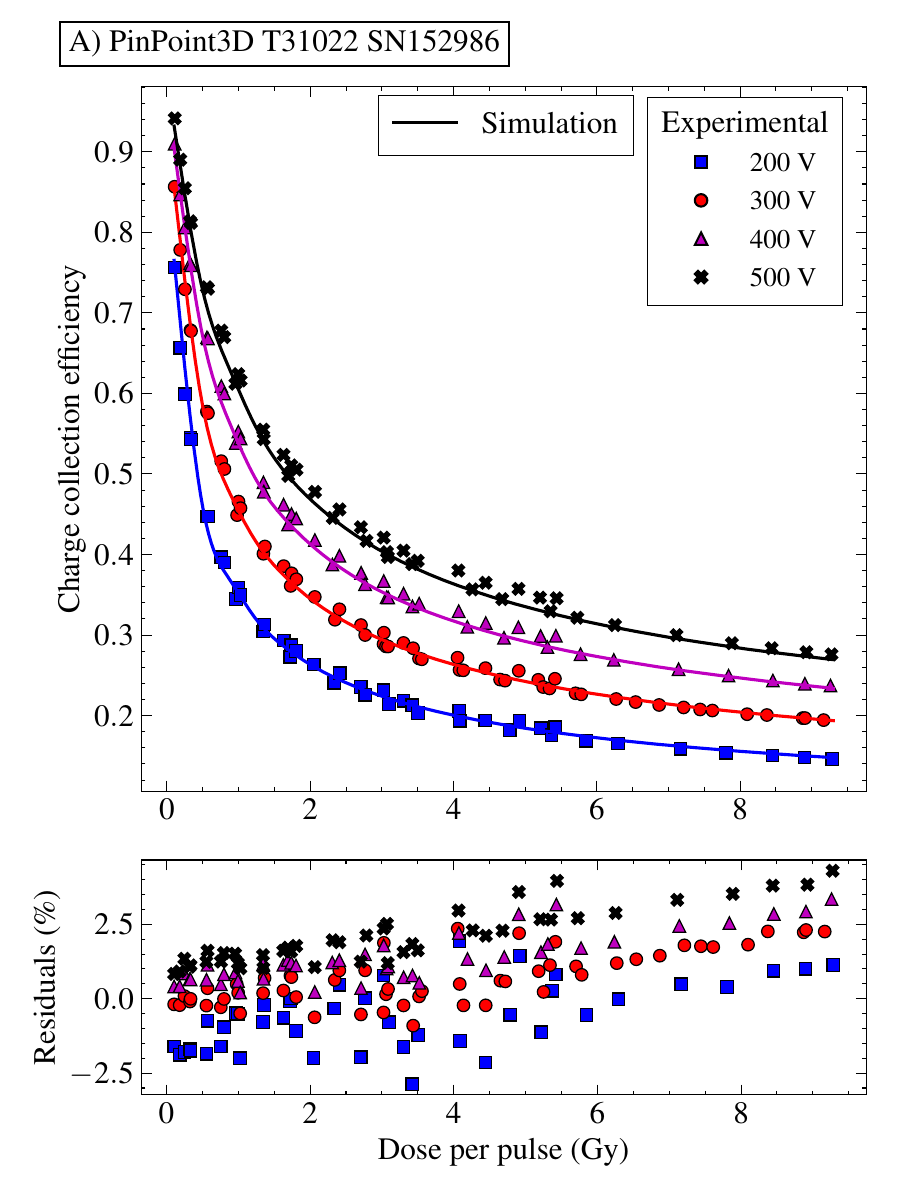}
\includegraphics[width = 6 cm]{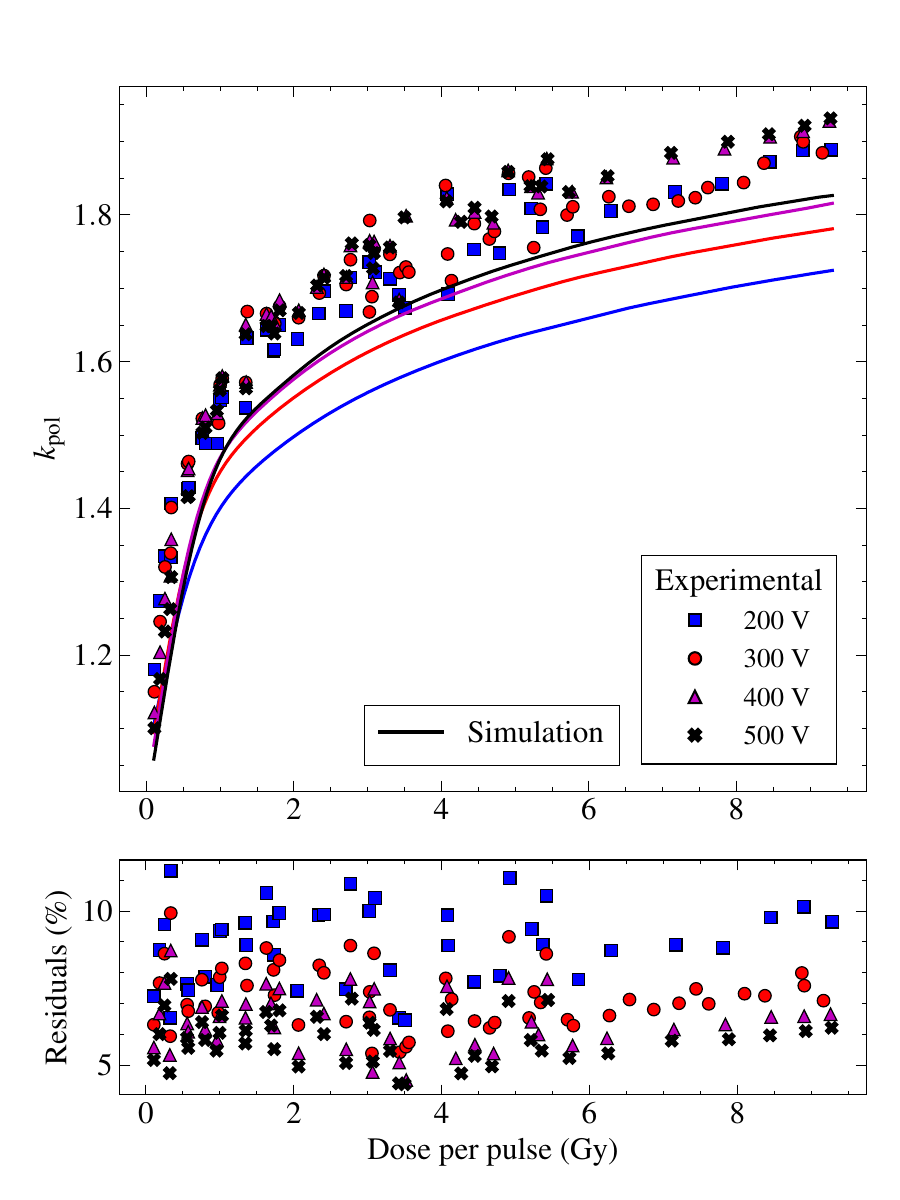}
\includegraphics[width = 6 cm]{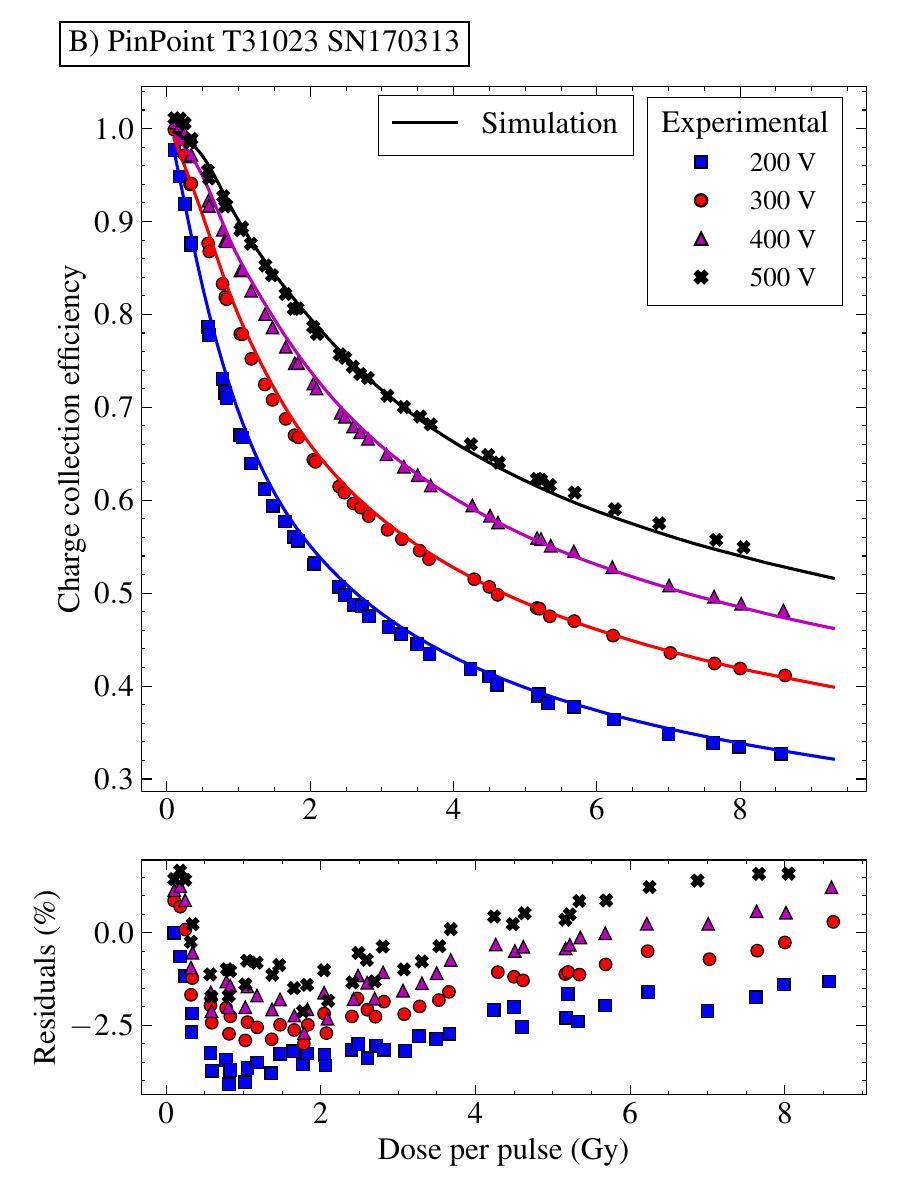}
\includegraphics[width = 6 cm]{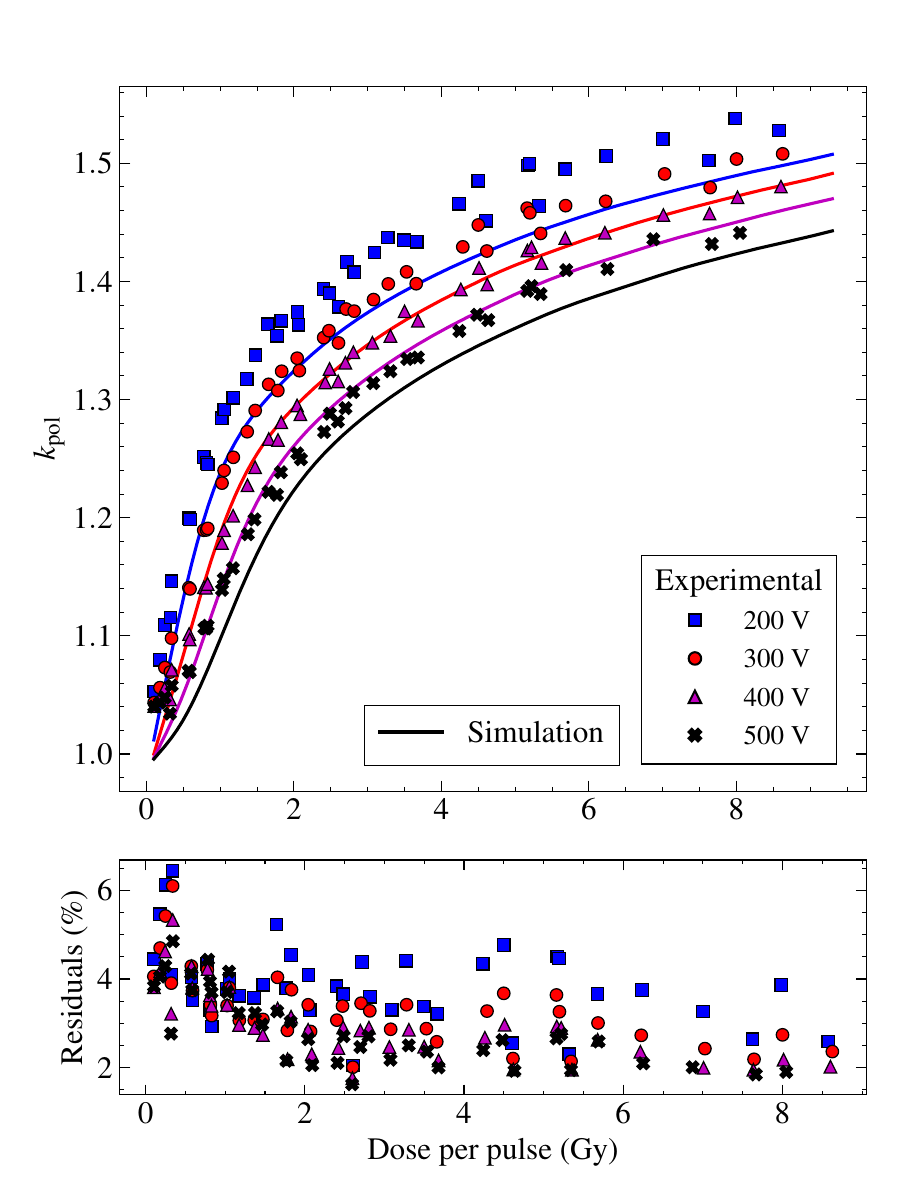}
\caption{Comparison between experimental and simulated CCE (left side) and polarity effect correction factor (right side) using the complete 2D geometry for the PinPoint3D T31022 SN 152986 (top, panel A) and the PinPoint T31023 SN 170313 IC (bottom, panel B) as a function of the DPP for different bias voltages. The pulse duration is 1.9~$\upmu$s. The results from simulations are shown using an idealized continuous line while the detailed residuals with respect to the experimental data are reported in the lower panels. Simulations were performed using ion mobilities from Boissonnat \textit{et al.}~\citep{Boissonnat_Measurement_2016} together with a volume recombination coefficient of 1.4$\times$10$^{-12}$~m$^3$~s$^{-1}$.}
\vspace{0.5 cm}
\label{Fig4}
\end{figure}

The experimental and simulated CCE and $k_{\text{pol}}$ are shown in Figure~\ref{Fig4} for the PP3D IC (SN~152986) and the PP IC (SN~170313) as a function of the DPP. These simulations were performed using Boissonnat \textit{et al.}~\citep{Boissonnat_Measurement_2016} mobilities together with a volume recombination coefficient of 1.4$\times$10$^{-12}$~m$^3$~s$^{-1}$. As expected, the CCE is larger when the collected charge sign is negative, resulting in $k_{\text{pol}} > 1$. On average, discrepancies between experimental and simulated CCE ($k_{\text{pol}}$) are 1.4~\% (7.0~\%) and 1.6~\% (3.2~\%) for the PP3D and PP ICs, respectively.

In general, a satisfactory agreement is found in terms of CCE and the trend for $k_{\rm pol}$ is described correctly. A systematic underestimation of $k_{\text{pol}}$ is observed, especially for the PP3D chamber.\\

A similar agreement is found when using the ion mobilities reported by Zhang \textit{et al.}~\citep{Zhang_Prediction_2019} together with an ion-ion recombination coefficient of 0.8$\times$10$^{-12}$~m$^3$~s$^{-1}$. On average, discrepancies between experimental and simulated CCE ($k_{\text{pol}}$) are 2.0~\% (7.3~\%) and 1.7~\% (3.6~\%) for the PP3D and PP ICs, respectively.\\

\begin{figure}[!t]
\centering
\includegraphics[width = 13.5 cm]{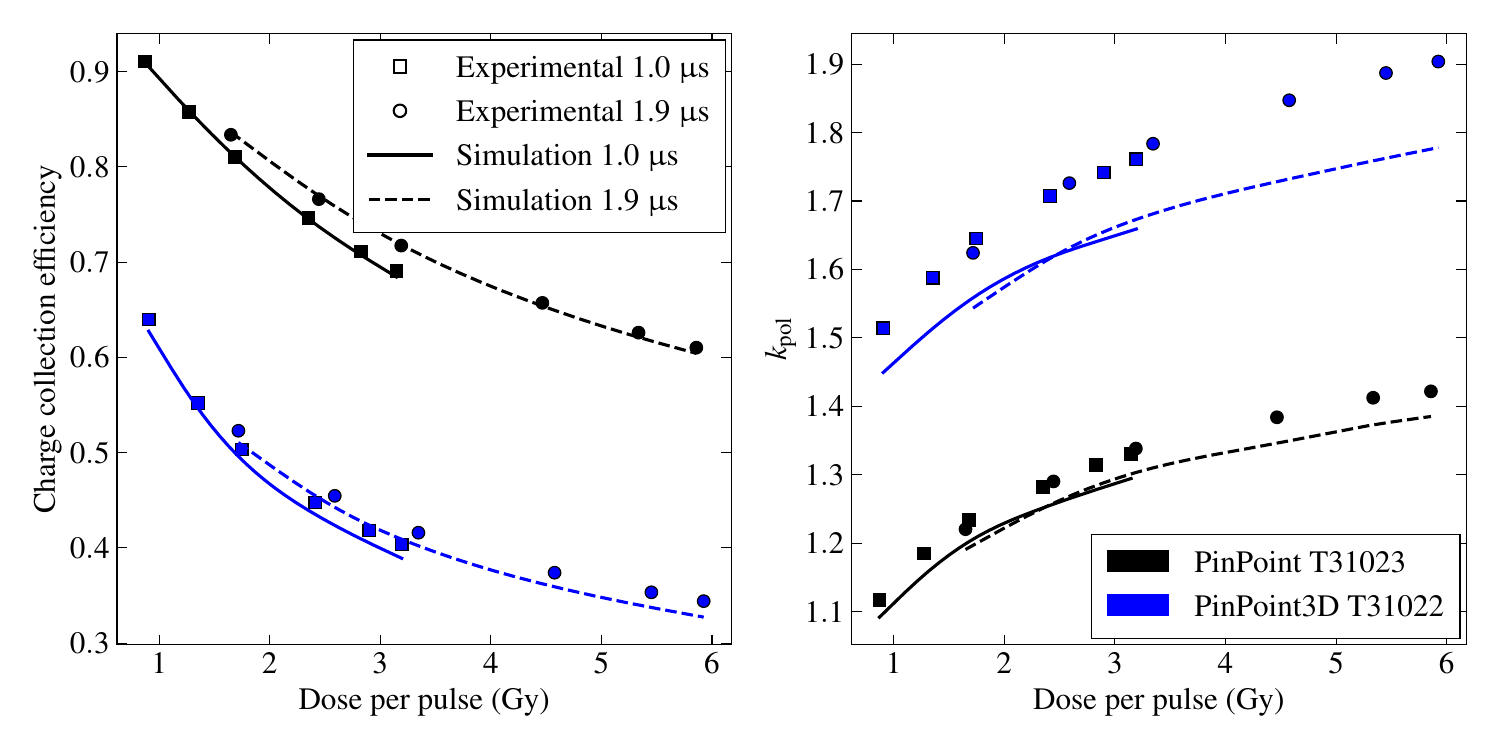}
\caption{Impact of the pulse duration on the CCE and $k_{\rm pol}$ for the PinPoint T31023 (SN 170293) and the PinPoint3D T31022 ICs (SN 153056) using 1.0~$\upmu$s and 1.9~$\upmu$s at 500~V bias voltage.}
\vspace{0.5 cm}
\label{pDuration}
\end{figure}

Regarding pulse duration, CCE increases as pulse duration becomes longer for a given DPP. Figure~\ref{pDuration} shows the influence of pulse duration on both CCE and $k_{\text{pol}}$ for the two ICs under a bias voltage of 500~V. In the overlapping DPP range (1.7~Gy to 3.1~Gy per pulse) differences in CCE up to 4.8~\% and 4.2~\% were found between 1.0~$\upmu$s and 1.9~$\upmu$s for the PP3D and PP ICs, respectively. The simulation successfully reproduces the observed trend and the relative variation in CCE due to pulse duration with discrepancies below 2~\% at 500~V.\\

The experimental difference in CCE ($k_{\text{pol}}$) on average for DPP from 1.6~Gy to 6~Gy between the two samples of each chamber type at 500~V was 4.5~\% (2.8~\%) and 1.4~\% (1.6~\%) for the PP3D and the PP ICs, respectively. This intra-model variability aligns with previous findings by Bourgouin \textit{et al.}~\citep{Bourgouin_Charge_2023} who reported variations up to 20~\% at 6~Gy per pulse in parallel-plate ICs.\\
\clearpage 

\clearpage

The use of a simplified 2D geometry leads to a systematic overestimation of CCE (see $f_{\rm 2D}$ and $f_{\rm 2D + gR}$ in Figure~\ref{Fig3}) with average discrepancies of 12~\% for the PP3D and 2.7~\% for the PP IC. Additionally, the simplified 2D model consistently overestimates the polarity correction factor, particularly at high DPP values with discrepancies respect to the experimental data reaching up to 10~\% for the PP3D. For the PP IC a slightly better agreement is found with average discrepancies for $k_{\text{pol}}$ around 2.8~\%.\\

\subsection{Time-resolved currents} 
\label{time_signal}

\begin{figure}[!t]
\centering
\includegraphics[width = 13.5 cm]{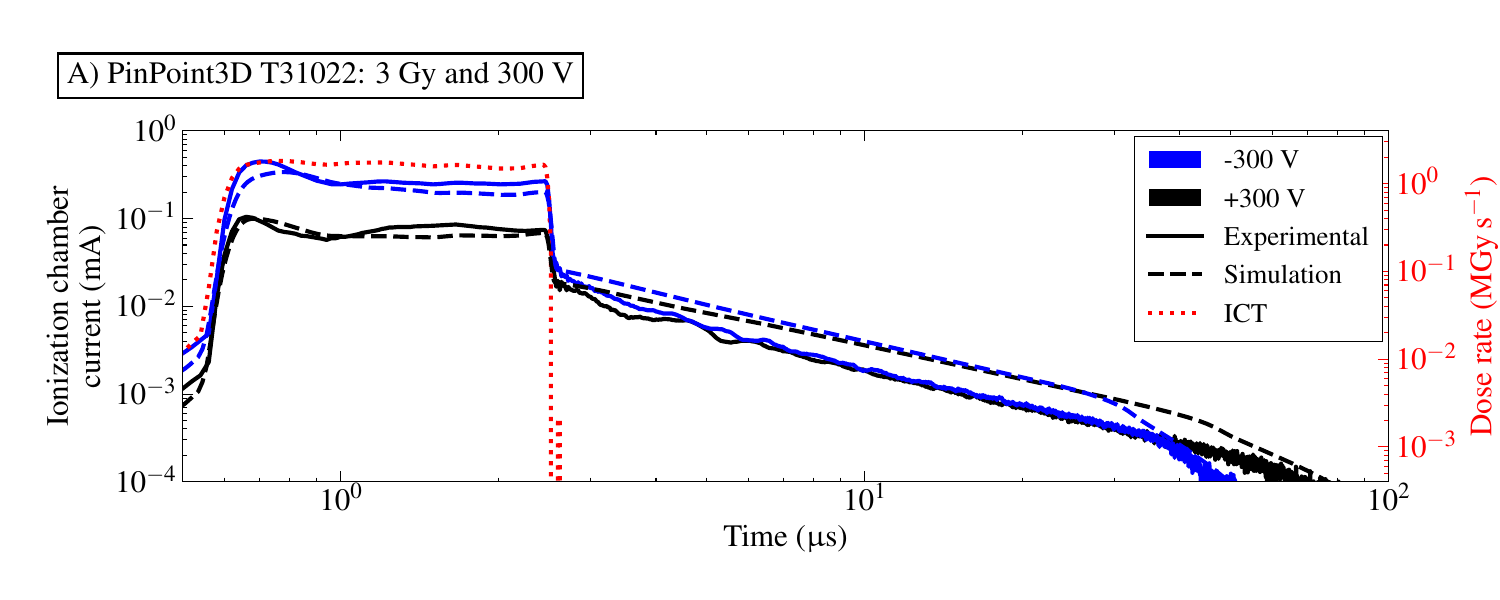}
\includegraphics[width = 13.5 cm]{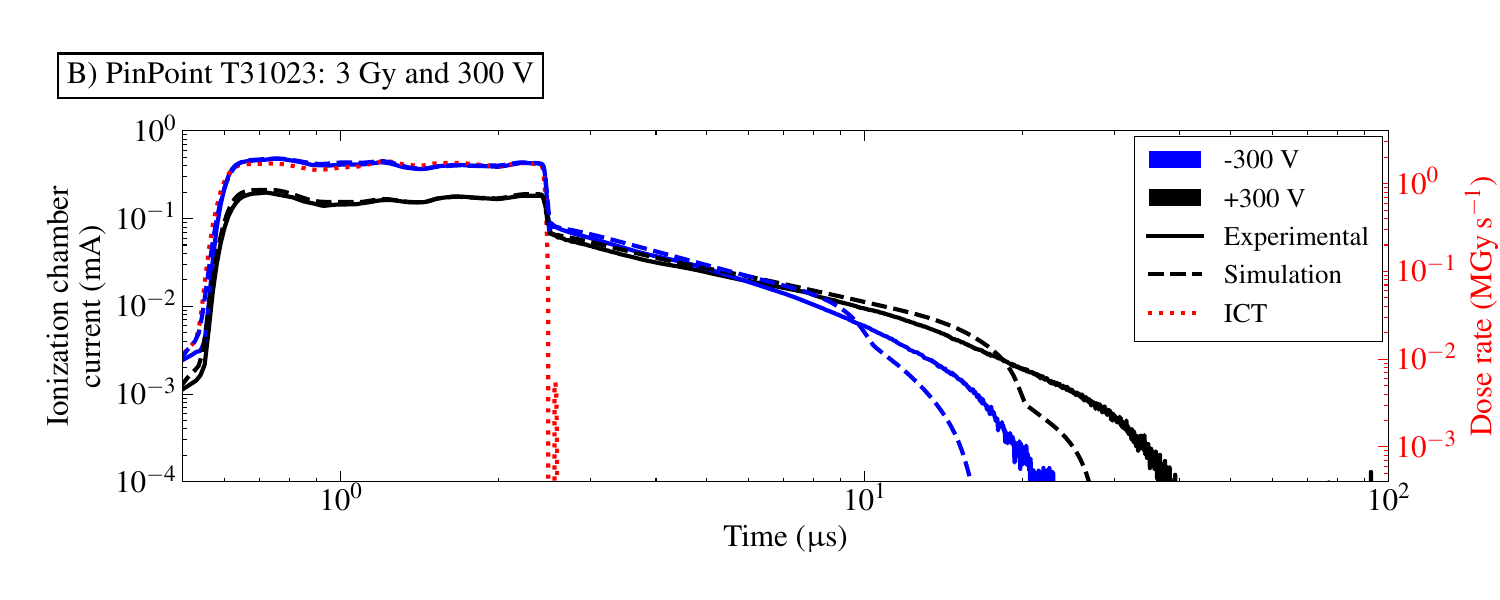}
\caption{Time-resolved signal obtained using a PinPoint 3D T31022 (panel A) and a PinPoint T31023 IC (panel B) IC irradiated with approximately 3 Gy per pulse in 1.9~$\upmu$s and with a bias voltage of +300~V (black) and -300~V (blue). The beam current used as an input for the simulation is shown in red, whose magnitude is displayed in the red right axis as dose rate. Numerical simulations use the Boissonnat \textit{et al.}~\citep{Boissonnat_Measurement_2016} ion mobilities.}
\label{I_IC}
\end{figure}

Similar to the behavior observed in parallel-plate ICs~\citep{Paz-Martin_Numerical_2022}, the time-resolved current of thimble ICs can be separated into two distinct contributions. During the pulse, a large current is generated by the rapid drift of free electrons within the IC. Variations in the pulse delivery time profile on the nanosecond scale significantly affect the IC’s instantaneous current due to the high mobility of electrons. After the beam pulse ends, a prolonged ion tail is observed, resulting from the slower drift of ion species. Figure~\ref{I_IC} shows the time-resolved currents for the PP3D (panel A) and the PP IC (panel B) for +300~V and -300~V bias voltage irradiated with a DPP of 3~Gy and a pulse duration of 1.9~$\upmu$s. Both chambers exhibit higher instantaneous current during the pulse at negative voltage. This corresponds to electrons drifting toward the central electrode, where the electric field is stronger. Additionally, charge collection is faster in this configuration, due to the combined effects of electric field perturbation and the slower drift of positive ions toward regions with lower field strength. When comparing both chambers under the same bias voltage, the PP chamber shows a faster charge collection time than the PP3D, primarily due to its smaller external radius.

\subsection[Invariant rules for the CCE for ICs]{Invariant rules for the charge collection efficiency of ionization chambers}
\label{invariant_rules}

Previous work of Kranzer \textit{et al.}~\citep{Kranzer_Charge_2022} showed that the CCE of two parallel-plate ICs $f_{1}$ and $f_{2}$ with distance between electrodes $d_{1}$ and $d_{2}$ biased with voltages $U_{1}$ and $U_{2} = {d_2}^2\,U_1/{d_1}^2$ are approximately equal under same irradiation conditions. This was verified both experimentally and with numerical simulation, giving a scaling rule as:
\begin{equation}
f_1\left(d_1, U_1\right)\approx f_2\left(d_2,\frac{{d_2}^2}{{d_1}^2}U_1\right)
\end{equation}

Bourgouin \textit{et al.}~\citep{Bourgouin_Charge_2023} evaluated this relationship across a broader set of parallel-plate ICs, demonstrating reasonable agreement among various ICs models. This invariant rule seems to be connected to the Boag model, where the CCE is a function of the dimensionless parameter $u$ (Equation~\ref{fenwick}), preserving this functional dependency. However, the applicability of the Boag model is restricted to DPPs values where the electric field perturbation can be neglected. Furthermore, even in the low DPP regime, its validity for air-vented ICs is compromised as it disregards the contribution of free electrons that drift across the IC without undergoing attachment~\citep{Paz-Martin_Evaluation_2025}.\\

Surprisingly, this simple invariant rule seems to work for air-vented parallel-plate ICs at large DPPs, where electric field perturbation largely impacts the charge transport. Recently, Fenwick \textit{et al.}~\citep{Fenwick_Collection_2024} have shown that the CCE of a cylindrical IC can be related to the CCE of a parallel-plate IC at the same bias voltage if the geometrical parameters $g_{\rm pp}$ and $g_{\rm cyl}$ are the same when the electric field perturbation is negligible.\\

\begin{figure}[!t]
\centering
\includegraphics[width = 8 cm]{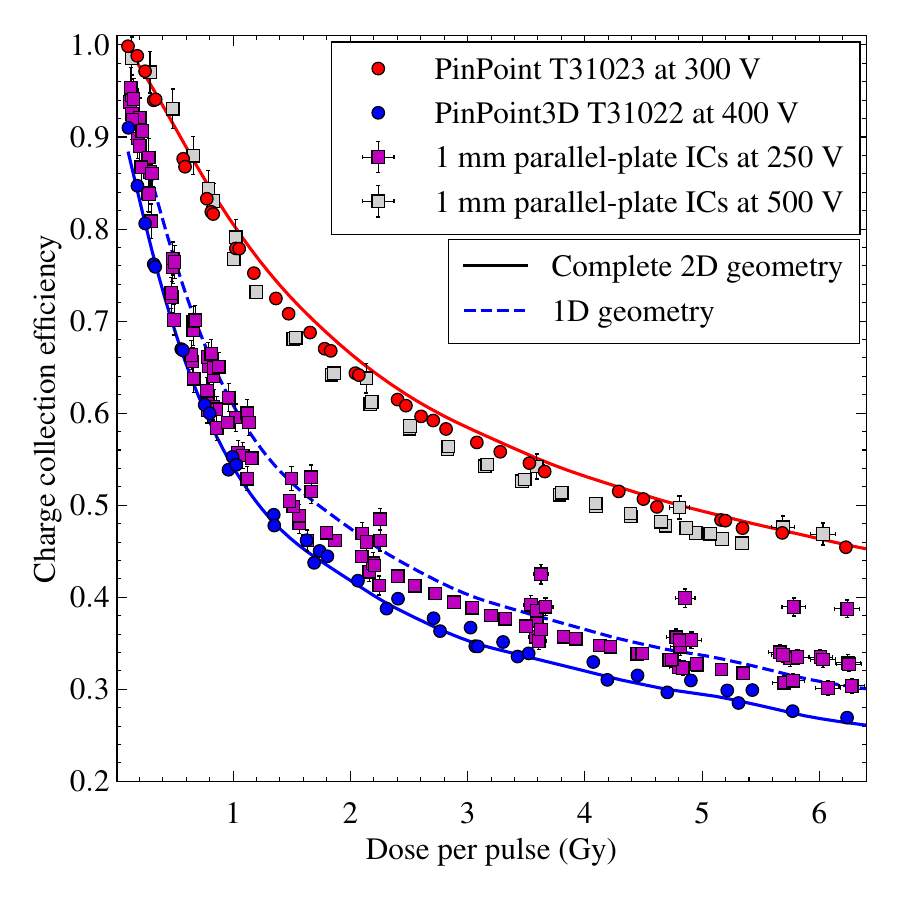}
\caption{CCE of the PinPoint 3D T31022 ($g_{\text{cyl}}/U = 3.96$~mm$^2$/kV) at 400~V and the PinPoint T31023 at 300~V  ($g_{\rm pp}/U = 1.83~$mm$^2$/kV) and their corresponding simulations versus the CCE of several 1~mm parallel-plate ICs operated at 250~V ($g_{\text{pp}}/U = 4.00~$mm$^2$/kV) and 500~V ($g_{\text{pp}}/U = 2.00~$mm$^2$/kV) from Bourgouin \textit{et al.}~\citep{Bourgouin_Charge_2023} (uncertainty corresponds to $k = 2$) and Kranzer \textit{et al.}~\citep{Kranzer_Charge_2022}. Chambers with comparable geometrical parameter over the operational bias voltage ($g/U$) shows similar charge collection efficiency.}
\label{scaling}
\end{figure}

Figure~\ref{scaling} displays the CCE measurements of several samples of Advanced Markus parallel-plate ICs (nominal distance between electrodes of 1~mm) from Bourgouin \textit{et al.}~\citep{Bourgouin_Charge_2023} and two 1~mm parallel-plate prototypes from Kranzer \textit{et al.}~\citep{Kranzer_Charge_2022}, using 250~V and 500~V bias voltages. The CCE of the PP3D operated at 400~V ($g_{\text{cyl}}/U = 3.96$~mm$^2$/kV) closely follows the trend observed for the parallel-plate ICs at 250~V ($g_{\text{pp}}/U = 4.00~$mm$^2$/kV). Similarly, the PP IC operated at 300~V ($g_{\rm pp}/U = 1.83~$mm$^2$/kV) shows a similar trend to that of the parallel-plate ICs at 500~V ($g_{\text{pp}}/U = 2.00~$mm$^2$/kV). For the PP3D, a slight deviation at large DPPs values might be attributed to the influence of the guard ring and the spherical tip on the CCE. In fact, the simulation using the 1D geometry yields a better agreement with the CCE of the parallel-plate ICs at 250~V. Here, the 1D simulation was performed using the real IC volume, in order to obtain a similar released charge in the chamber for a given DPP as in the parallel-plate ICs.

\section{Discussion}

According to the results presented in section \ref{weighted_CCE}, the CCE of a thimble IC can be reasonably approximated by the volume-weighted average CCE of simple spherical and cylindrical geometries if no guard ring is present. The discrepancy between the weighted approach and the full simulation is surprisingly smaller than 3~\% whenever the guard ring does not substantially affect the chamber response. The presence of a guard ring electrode introduces an insensitive region within the chamber (dead volume) and affects charge transport, leading to a reduction in both polarity correction factor and CCE. Therefore, in order to accurately reproduce the CCE of thimble ICs, especially when the dead volume is an important fraction of the total volume (nearly 30~\% for the PP3D compared to 6~\% for the PP), it is necessary to include the guard ring in the simulation. This dead volume also explains the small difference observed in the calibration coefficients between the two ICs types of about +4.6~\% experimentally and +5.0~\% predicted by simulation, while the physical volume difference between the ICs is about -15~\%. \\

As shown in Section~\ref{CCE_kpol}, both experimental data and simulations indicate that the studied ICs exhibit an increasing polarity effect correction factor, $k_{\rm pol}$, with increasing DPP. The numerical model attributes a large fraction of this polarity effect to volume recombination processes that depend on the polarity of the applied bias voltage, contrary to low DPP where volume recombination no longer plays a major role and the polarity effect is primarily caused by extracavitary charge contributions. Consequently, for thimble-type chambers operating at high DPP, there is an interplay between the CCE and the polarity correction factor, making their separation and interpretation more complex.
The discrepancies observed in terms of $k_{\text{pol}}$ might indicate limitations
in the implementation of the modeled IC geometry. However, as reported previously by Bourgouin \textit{et al.}~\citep{Bourgouin_Charge_2023}, parallel-plate ICs irradiated with large DPPs also exhibit an increasing polarity effect with DPP. In addition to polarity dependent charge transport and recombination, other factors may contribute to the observed polarity effect. In order to understand the mechanisms leading to the polarity effect observed in IC exposed to ultra-high-DPP further research is required. \\

The invariant rules used to relate the CCE of different ICs, as discussed in Section~\ref{invariant_rules}, were previously identified by Kranzer \textit{et al.}~\citep{Kranzer_Charge_2022} for parallel-plate ICs. The use of the parameter $g/U$, with units of length squared over bias voltage, appears to provide an invariant for CCE, with a rough agreement of approximately 5~\%. Extending this approach to cylindrical chambers has demonstrated its usefulness not only for comparing different thimble-type ICs, but also for establishing relationships between parallel-plate and thimble chambers. However, since the parameter $g_{\rm cyl}$ is derived from an idealized cylindrical geometry, the resulting scaling rule should be expected to provide only an approximate prediction for chamber intercomparison.\\

Uncertainties remain in the transport parameters of charge carriers in air generated by ionizing radiation, which affect the accuracy of numerical models. To describe CCE and the observed polarity correction factor $k_{\text{pol}}$, the use of different average ion mobilities has a limited impact, provided that the ion-ion recombination coefficient is properly adjusted.

\section{Conclusions}

The experimental results show that CCE is higher when collected charge sign is negative, which contributes to a polarity correction factor $k_{\text{pol}} > 1$ attributed to the polarity-dependent charge transport and recombination within the chamber. At large DPPs, where the polarity correction factor can reach values up to 1.9, thimble ICs should be used with caution due to their large polarization effect. CCE was found to be dependent on the pulse duration, with relative differences up to 4.8~\% between 1.0~$\upmu$s and 1.9~$\upmu$s for 500~V. Although the PP and the PP3D IC have a similar sensitive volume, the PP shows a larger CCE for the same value of DPP and bias voltage compared to the PP3D, explained by the reduced outer radius electrode.\\

The experimental data have been compared to a novel finite-element based numerical model that enables simulation of the detailed 2D geometry of the chambers, assuming rotational symmetry around the central axis. This model successfully reproduces the polarity-dependent charge collection behavior, which arises from  different effective volume recombination in the chamber for each polarity. Consequently, under high DPPs conditions, thimble-type chambers exhibit an interplay between recombination effects and polarity effect, complicating the interpretation of the measured response. Based on numerical simulations, the CCE of a thimble IC can be described as the volume-weighted average of a cylindrical and a spherical geometry. Discrepancies between this weighted model and the simplified 2D model below 2.5~\% and 0.2~\% are observed for chambers with the geometries of the PP3D and PP, respectively. The differences between CCE simulated taking into account the guard ring versus a simplified 2D model can reach up to 18~\% for the PP3D and 1.7~\% for the PP. Thus, the use of a complete 2D model is mandatory to accurately describe the CCE and $k_{\rm pol}$ of the PP3D.\\

Using the complete 2D geometry, deviations between experimental and simulated CCE ($k_{\text{pol}}$) are 1.4~\% (7.0~\%) and 1.6~\% (3.2~\%) for the PP3D and PP ICs, respectively. Overall, the numerical model shows a satisfactory agreement in terms of CCE and a correct description of the observed trend for $k_{\text{pol}}$.\\

Finally, the CCE of the PP and PP3D ICs can be related to the CCE of a 1~mm parallel-plate IC using the ratio of the \emph{so-called} geometrical parameter and bias voltage $g/U$. Deviations at large DPP from this tendency for the PP3D might be attributed to the impact of the guard ring electrode and the tip of the IC. This relation is useful to compare the response of different ICs types under ultra-high DPPs.

\section*{Acknowledgments}

\noindent José Paz-Martín has received a predoctoral research contract from the Xunta de Galicia regional government. Juan Pardo-Montero acknowledges the support of Xunta de Galicia, Axencia Galega de Innovación (grant IN607D2022/02). This work has received funding from `La Caixa' Foundation under the grant agreement HR23-00718 (`Dosimetry monitor for FLASH therapy' project), and from grant PLEC2022-009476 funded by MCIN/AEI/10.13039/501100011033. The project (24NRM01 FLASH-DOSE) has received funding from the European Partnership on Metrology, co-financed from the European Union's Horizon Europe Research and Innovation Programme and by the Participating States. The authors thank Christoph Makowski for maintaining the electron linear accelerator during the measurement campaign, Olaf Tappe for the machining work, and Thomas Hackel for the alanine measurements. We also thank Rafael Kranzer (PTW) for kindly providing the samples of the ICs used in the experimental characterization and for supplying the blueprints for the numerical and Monte-Carlo simulations.

\section*{Data availability}
The data supporting these findings is available at the following repository: \url{https://doi.org/10.5281/zenodo.17737817}. The numerical code is available at \url{https://github.com/JPazMartin/ICSimFEM}.

\section*{Conflict of Interest}
The authors have no relevant conflicts of interest to disclose.

\end{document}